\documentclass[12pt,a4paper]{article}
\makeatletter
\newcount\@hour\newcount\@minute\chardef\@x10\chardef\@xv60
\def\tcitime{
\def\@time{%
  \@minute\time\@hour\@minute\divide\@hour\@xv
  \ifnum\@hour<\@x 0\fi\the\@hour:%
  \multiply\@hour\@xv\advance\@minute-\@hour
  \ifnum\@minute<\@x 0\fi\the\@minute
  }}%
\def\QCTOpt[#1]#2{%
  \def\QCTOptB{#1}
  \def\QCTOptA{#2}
}
\def\QCTNOpt#1{%
  \def\QCTOptA{#1}
  \let\QCTOptB\empty
}
\def\Qct{%
  \@ifnextchar[{%
    \QCTOpt}{\QCTNOpt}
}
\def\QCBOpt[#1]#2{%
  \def\QCBOptB{#1}
  \def\QCBOptA{#2}
}
\def\QCBNOpt#1{%
  \def\QCBOptA{#1}
  \let\QCBOptB\empty
}
\def\Qcb{%
  \@ifnextchar[{%
    \QCBOpt}{\QCBNOpt}
}
\def\PrepCapArgs{%
  \ifx\QCBOptA\empty
    \ifx\QCTOptA\empty
      {}%
    \else
      \ifx\QCTOptB\empty
        {\QCTOptA}%
      \else
        [\QCTOptB]{\QCTOptA}%
      \fi
    \fi
  \else
    \ifx\QCBOptA\empty
      {}%
    \else
      \ifx\QCBOptB\empty
        {\QCBOptA}%
      \else
        [\QCBOptB]{\QCBOptA}%
      \fi
    \fi
  \fi
}
\newcount\GRAPHICSTYPE
\GRAPHICSTYPE=\z@
\def\GRAPHICSPS#1{%
 \ifcase\GRAPHICSTYPE
   \special{ps: #1}%
 \or
   \special{language "PS", include "#1"}%
 \fi
}%
\def\graffile#1#2#3#4{%
    \leavevmode
    \raise -#4 \BOXTHEFRAME{%
        \hbox to #2{\raise #3\hbox{\null #1}}}%
}%
\def\draftbox#1#2#3#4{%
 \leavevmode\raise -#4 \hbox{%
  \frame{\rlap{\protect\tiny #1}\hbox to #2%
   {\vrule height#3 width\z@ depth\z@\hfil}%
  }%
 }%
}%
\newcount\draft
\draft=\z@

\newif\ifwasdraft
\wasdraftfalse

\def\GRAPHIC#1#2#3#4#5{%
 \ifnum\draft=\@ne\draftbox{#2}{#3}{#4}{#5}%
  \else\graffile{#1}{#3}{#4}{#5}%
  \fi
 }%
\def\addtoLaTeXparams#1{%
    \edef\LaTeXparams{\LaTeXparams #1}}%
\newif\ifBoxFrame \BoxFramefalse
\newif\ifOverFrame \OverFramefalse
\newif\ifUnderFrame \UnderFramefalse

\def\BOXTHEFRAME#1{%
   \hbox{%
      \ifBoxFrame
         \frame{#1}%
      \else
         {#1}%
      \fi
   }%
}

\def\doFRAMEparams#1{\BoxFramefalse\OverFramefalse\UnderFramefalse\readFRAMEparams#1\end}%
\def\readFRAMEparams#1{%
 \ifx#1\end%
  \let\next=\relax
  \else
  \ifx#1i\dispkind=\z@\fi
  \ifx#1d\dispkind=\@ne\fi
  \ifx#1f\dispkind=\tw@\fi
  \ifx#1t\addtoLaTeXparams{t}\fi
  \ifx#1b\addtoLaTeXparams{b}\fi
  \ifx#1p\addtoLaTeXparams{p}\fi
  \ifx#1h\addtoLaTeXparams{h}\fi
  \ifx#1X\BoxFrametrue\fi
  \ifx#1O\OverFrametrue\fi
  \ifx#1U\UnderFrametrue\fi
  \ifx#1w
    \ifnum\draft=1\wasdrafttrue\else\wasdraftfalse\fi
    \draft=\@ne
  \fi
  \let\next=\readFRAMEparams
  \fi
 \next
 }%
\def\IFRAME#1#2#3#4#5#6{%
      \bgroup
      \let\QCTOptA\empty
      \let\QCTOptB\empty
      \let\QCBOptA\empty
      \let\QCBOptB\empty
      #6%
      \parindent=0pt%
      \leftskip=0pt
      \rightskip=0pt
      \setbox0 = \hbox{\QCBOptA}%
      \@tempdima = #1\relax
      \ifOverFrame
          \typeout{This is not implemented yet}%
          \show\HELP
      \else
         \ifdim\wd0>\@tempdima
            \advance\@tempdima by \@tempdima
            \ifdim\wd0 >\@tempdima
               \textwidth=\@tempdima
               \setbox1 =\vbox{%
                  \noindent\hbox to \@tempdima{\hfill\GRAPHIC{#5}{#4}{#1}{#2}{#3}\hfill}\\%
                  \noindent\hbox to \@tempdima{\parbox[b]{\@tempdima}{\QCBOptA}}%
               }%
               \wd1=\@tempdima
            \else
               \textwidth=\wd0
               \setbox1 =\vbox{%
                 \noindent\hbox to \wd0{\hfill\GRAPHIC{#5}{#4}{#1}{#2}{#3}\hfill}\\%
                 \noindent\hbox{\QCBOptA}%
               }%
               \wd1=\wd0
            \fi
         \else
            \ifdim\wd0>0pt
              \hsize=\@tempdima
              \setbox1 =\vbox{%
                \unskip\GRAPHIC{#5}{#4}{#1}{#2}{0pt}%
                \break
                \unskip\hbox to \@tempdima{\hfill \QCBOptA\hfill}%
              }%
              \wd1=\@tempdima
           \else
              \hsize=\@tempdima
              \setbox1 =\vbox{%
                \unskip\GRAPHIC{#5}{#4}{#1}{#2}{0pt}%
              }%
              \wd1=\@tempdima
           \fi
         \fi
         \@tempdimb=\ht1
         \advance\@tempdimb by \dp1
         \advance\@tempdimb by -#2%
         \advance\@tempdimb by #3%
         \leavevmode
         \raise -\@tempdimb \hbox{\box1}%
      \fi
      \egroup%
}%
\def\DFRAME#1#2#3#4#5{%
 \begin{center}
     \let\QCTOptA\empty
     \let\QCTOptB\empty
     \let\QCBOptA\empty
     \let\QCBOptB\empty
     \ifOverFrame 
        #5\QCTOptA\par
     \fi
     \GRAPHIC{#4}{#3}{#1}{#2}{\z@}
     \ifUnderFrame 
        \par #5\QCBOptA
     \fi
 \end{center}%
 }%
\def\FFRAME#1#2#3#4#5#6#7{%
 \begin{figure}[#1]%
  \let\QCTOptA\empty
  \let\QCTOptB\empty
  \let\QCBOptA\empty
  \let\QCBOptB\empty
  \ifOverFrame
    #4
    \ifx\QCTOptA\empty
    \else
      \ifx\QCTOptB\empty
        \caption{\QCTOptA}%
      \else
        \caption[\QCTOptB]{\QCTOptA}%
      \fi
    \fi
    \ifUnderFrame\else
      \label{#5}%
    \fi
  \else
    \UnderFrametrue%
  \fi
  \begin{center}\GRAPHIC{#7}{#6}{#2}{#3}{\z@}\end{center}%
  \ifUnderFrame
    #4
    \ifx\QCBOptA\empty
      \caption{}%
    \else
      \ifx\QCBOptB\empty
        \caption{\QCBOptA}%
      \else
        \caption[\QCBOptB]{\QCBOptA}%
      \fi
    \fi
    \label{#5}%
  \fi
  \end{figure}%
 }%
\newcount\dispkind%
\def\FRAME#1#2#3#4#5#6#7#8{%
 \ifnum\draft=\@ne
   \wasdrafttrue
 \else
   \wasdraftfalse%
 \fi
 \def\LaTeXparams{}%
 \dispkind=\z@
 \def\LaTeXparams{}%
 \doFRAMEparams{#1}%
 \ifnum\dispkind=\z@\IFRAME{#2}{#3}{#4}{#7}{#8}{#5}\else
  \ifnum\dispkind=\@ne\DFRAME{#2}{#3}{#7}{#8}{#5}\else
   \ifnum\dispkind=\tw@
    \edef\@tempa{\noexpand\FFRAME{\LaTeXparams}}%
    \@tempa{#2}{#3}{#5}{#6}{#7}{#8}%
    \fi
   \fi
  \fi
  \ifwasdraft\draft=1\else\draft=0\fi{}%
 }%
\def\TEXUX#1{"texux"}

\def\limfunc#1{\mathop{\rm #1}}%

\long\def\QQQ#1#2{%
     \long\expandafter\def\csname#1\endcsname{#2}}%
\@ifundefined{QTP}{\def\QTP#1{}}{}
\@ifundefined{Qlb}{}{}
\@ifundefined{Qlt}{}{}
\long\def\QQA#1#2{}%
\def\QTR#1#2{{\csname#1\endcsname #2}}
\def\EXPAND#1[#2]#3{}%
\def\NOEXPAND#1[#2]#3{}%
\def\LaTeXparent#1{}%
\def\ChildStyles#1{}%
\def\ChildDefaults#1{}%
\def\QTagDef#1#2#3{}%
\def\QQfnmark#1{\footnotemark}

\@ifundefined{INDEX}{\def\INDEX#1#2{}{}}{}%
\@ifundefined{SUBINDEX}{\def\SUBINDEX#1#2#3{}{}{}}{}%
\def\initial#1{\bigbreak{\raggedright\large\bf #1}\kern 2\p@
   \penalty3000}%
\@ifundefined{ZZZ}{}{\makeatletter\input gnuindex.sty\makeatother\makeindex\makeatletter}%
\@ifundefined{abstract}{%
 \def\abstract{%
  \if@twocolumn
   \section*{Abstract (Not appropriate in this style!)}%
   \else \small 
   \begin{center}{\bf Abstract\vspace{-.5em}\vspace{\z@}}\end{center}%
   \quotation 
   \fi
  }%
 }{%
 }%
\@ifundefined{endabstract}{\def\endabstract
  {\if@twocolumn\else\endquotation\fi}}{}%
\@ifundefined{maketitle}{\def\maketitle#1{}}{}%
\@ifundefined{affiliation}{\def\affiliation#1{}}{}%
\@ifundefined{proof}{}{}%
\@ifundefined{endproof}{}{}%
\@ifundefined{newfield}{\def\newfield#1#2{}}{}%
\@ifundefined{chapter}{\def\chapter#1{\par(Chapter head:)#1\par }%
 \newcount\c@chapter}{}%
\@ifundefined{part}{\def\part#1{\par(Part head:)#1\par }}{}%
\@ifundefined{section}{\def\section#1{\par(Section head:)#1\par }}{}%
\@ifundefined{subsection}{\def\subsection#1%
 {\par(Subsection head:)#1\par }}{}%
\@ifundefined{subsubsection}{\def\subsubsection#1%
 {\par(Subsubsection head:)#1\par }}{}%
\@ifundefined{paragraph}{\def\paragraph#1%
 {\par(Subsubsubsection head:)#1\par }}{}%
\@ifundefined{subparagraph}{\def\subparagraph#1%
 {\par(Subsubsubsubsection head:)#1\par }}{}%
\@ifundefined{therefore}{}{}%
\@ifundefined{backepsilon}{}{}%
\@ifundefined{yen}{}{}%
\@ifundefined{registered}{%
   \def\registered{\relax\ifmmode{}\r@gistered
                    \else$\m@th\r@gistered$\fi}%
 \def\r@gistered{^{\ooalign
  {\hfil\raise.07ex\hbox{$\scriptstyle\rm\text{R}$}\hfil\crcr
  \mathhexbox20D}}}}{}%
\@ifundefined{Eth}{}{}%
\@ifundefined{eth}{}{}%
\@ifundefined{Thorn}{}{}%
\@ifundefined{thorn}{}{}%
\@ifundefined{degree}{}{}%
\newdimen\theight
\def\Column{%
 \vadjust{\setbox\z@=\hbox{\scriptsize\quad\quad tcol}%
  \theight=\ht\z@\advance\theight by \dp\z@\advance\theight by \lineskip
  \kern -\theight \vbox to \theight{%
   \rightline{\rlap{\box\z@}}%
   \vss
   }%
  }%
 }%
\def\qed{%
 \ifhmode\unskip\nobreak\fi\ifmmode\ifinner\else\hskip5\p@\fi\fi
 \hbox{\hskip5\p@\vrule width4\p@ height6\p@ depth1.5\p@\hskip\p@}%
 }%
\def\miss{\hbox{\vrule height2\p@ width 2\p@ depth\z@}}%
\def\tcol#1{{\baselineskip=6\p@ \vcenter{#1}} \Column}  %
\def\newfmtname{LaTeX2e}
\def\chkcompat{%
   \if@compatibility
   \else
     \usepackage{latexsym}
   \fi
}

\ifx\fmtname\newfmtname
  \DeclareOldFontCommand{\rm}{\normalfont\rmfamily}{\mathrm}
  \DeclareOldFontCommand{\sf}{\normalfont\sffamily}{\mathsf}
  \DeclareOldFontCommand{\tt}{\normalfont\ttfamily}{\mathtt}
  \DeclareOldFontCommand{\bf}{\normalfont\bfseries}{\mathbf}
  \DeclareOldFontCommand{\it}{\normalfont\itshape}{\mathit}
  \DeclareOldFontCommand{\sl}{\normalfont\slshape}{\@nomath\sl}
  \DeclareOldFontCommand{\sc}{\normalfont\scshape}{\@nomath\sc}
  \chkcompat
\fi

\@ifundefined{theorem}{}{}
\@ifundefined{lemma}{}{}
\@ifundefined{corollary}{}{}
\@ifundefined{conjecture}{}{}
\@ifundefined{proposition}{}{}
\@ifundefined{axiom}{}{}
\@ifundefined{remark}{}{}
\@ifundefined{example}{}{}
\@ifundefined{exercise}{}{}
\@ifundefined{definition}{}{}

\@ifundefined{mathletters}{%
  \newcounter{equationnumber}  
  \def\mathletters{%
     \addtocounter{equation}{1}
     \edef\@currentlabel{\theequation}%
     \setcounter{equationnumber}{\c@equation}
     \setcounter{equation}{0}%
     \edef\theequation{\@currentlabel\noexpand\alph{equation}}%
  }
  
}{}

\@ifundefined{BibTeX}{%
    \def\BibTeX{{\rm B\kern-.05em{\sc i\kern-.025em b}\kern-.08em
                 T\kern-.1667em\lower.7ex\hbox{E}\kern-.125emX}}}{}%
\@ifundefined{AmS}%
    {\def\AmS{{\protect\usefont{OMS}{cmsy}{m}{n}%
                A\kern-.1667em\lower.5ex\hbox{M}\kern-.125emS}}}{}%
\@ifundefined{AmSTeX}{}{}%
\let\DOTSI\relax
\def\RIfM@{\relax\ifmmode}%
\def\FN@{\futurelet\next}%
\newcount\intno@
\def\iint{\DOTSI\intno@\tw@\FN@\ints@}%
\def\iiint{\DOTSI\intno@\thr@@\FN@\ints@}%
\def\iiiint{\DOTSI\intno@4 \FN@\ints@}%
\def\idotsint{\DOTSI\intno@\z@\FN@\ints@}%
\def\ints@{\findlimits@\ints@@}%
\newif\iflimtoken@
\newif\iflimits@
\def\findlimits@{\limtoken@true\ifx\next\limits\limits@true
 \else\ifx\next\nolimits\limits@false\else
 \limtoken@false\ifx\ilimits@\nolimits\limits@false\else
 \ifinner\limits@false\else\limits@true\fi\fi\fi\fi}%
\def\multint@{\int\ifnum\intno@=\z@\intdots@                          
 \else\intkern@\fi                                                    
 \ifnum\intno@>\tw@\int\intkern@\fi                                   
 \ifnum\intno@>\thr@@\int\intkern@\fi                                 
 \int}
\def\multintlimits@{\intop\ifnum\intno@=\z@\intdots@\else\intkern@\fi
 \ifnum\intno@>\tw@\intop\intkern@\fi
 \ifnum\intno@>\thr@@\intop\intkern@\fi\intop}%
\def\intic@{%
    \mathchoice{\hskip.5em}{\hskip.4em}{\hskip.4em}{\hskip.4em}}%
\def\negintic@{\mathchoice
 {\hskip-.5em}{\hskip-.4em}{\hskip-.4em}{\hskip-.4em}}%
\def\ints@@{\iflimtoken@                                              
 \def\ints@@@{\iflimits@\negintic@
   \mathop{\intic@\multintlimits@}\limits                             
  \else\multint@\nolimits\fi                                          
  \eat@}
 \else                                                                
 \def\ints@@@{\iflimits@\negintic@
  \mathop{\intic@\multintlimits@}\limits\else
  \multint@\nolimits\fi}\fi\ints@@@}%
\def\intkern@{\mathchoice{\!\!\!}{\!\!}{\!\!}{\!\!}}%
\def\plaincdots@{\mathinner{\cdotp\cdotp\cdotp}}%
\def\intdots@{\mathchoice{\plaincdots@}%
 {{\cdotp}\mkern1.5mu{\cdotp}\mkern1.5mu{\cdotp}}%
 {{\cdotp}\mkern1mu{\cdotp}\mkern1mu{\cdotp}}%
 {{\cdotp}\mkern1mu{\cdotp}\mkern1mu{\cdotp}}}%
\def\RIfM@{\relax\protect\ifmmode}
\def\text{\RIfM@\expandafter\text@\else\expandafter\mbox\fi}
\let\nfss@text\text
\def\text@#1{\mathchoice
   {\textdef@\displaystyle\f@size{#1}}%
   {\textdef@\textstyle\tf@size{\firstchoice@false #1}}%
   {\textdef@\textstyle\sf@size{\firstchoice@false #1}}%
   {\textdef@\textstyle \ssf@size{\firstchoice@false #1}}%
   \glb@settings}

\def\textdef@#1#2#3{\hbox{{%
                    \everymath{#1}%
                    \let\f@size#2\selectfont
                    #3}}}
\newif\iffirstchoice@
\firstchoice@true
\def\Let@{\relax\iffalse{\fi\let\\=\cr\iffalse}\fi}%
\def\vspace@{\def\vspace##1{\crcr\noalign{\vskip##1\relax}}}%
\def\multilimits@{\bgroup\vspace@\Let@
 \baselineskip\fontdimen10 \scriptfont\tw@
 \advance\baselineskip\fontdimen12 \scriptfont\tw@
 \lineskip\thr@@\fontdimen8 \scriptfont\thr@@
 \lineskiplimit\lineskip
 \vbox\bgroup\ialign\bgroup\hfil$\m@th\scriptstyle{##}$\hfil\crcr}%
\def\Sb{_\multilimits@}%
\def\endSb{\crcr\egroup\egroup\egroup}%
\def\Sp{^\multilimits@}%

\newdimen\ex@
\def\rightarrowfill@#1{$#1\m@th\mathord-\mkern-6mu\cleaders
 \hbox{$#1\mkern-2mu\mathord-\mkern-2mu$}\hfill
 \mkern-6mu\mathord\rightarrow$}%
\def\leftarrowfill@#1{$#1\m@th\mathord\leftarrow\mkern-6mu\cleaders
 \hbox{$#1\mkern-2mu\mathord-\mkern-2mu$}\hfill\mkern-6mu\mathord-$}%
\def\leftrightarrowfill@#1{$#1\m@th\mathord\leftarrow
\mkern-6mu\cleaders
 \hbox{$#1\mkern-2mu\mathord-\mkern-2mu$}\hfill
 \mkern-6mu\mathord\rightarrow$}%
\def\overrightarrow{\mathpalette\overrightarrow@}%
\def\overrightarrow@#1#2{\vbox{\ialign{##\crcr\rightarrowfill@#1\crcr
 \noalign{\kern-\ex@\nointerlineskip}$\m@th\hfil#1#2\hfil$\crcr}}}%

\def\overleftarrow{\mathpalette\overleftarrow@}%
\def\overleftarrow@#1#2{\vbox{\ialign{##\crcr\leftarrowfill@#1\crcr
 \noalign{\kern-\ex@\nointerlineskip}$\m@th\hfil#1#2\hfil$\crcr}}}%
\def\overleftrightarrow{\mathpalette\overleftrightarrow@}%
\def\overleftrightarrow@#1#2{\vbox{\ialign{##\crcr
   \leftrightarrowfill@#1\crcr
 \noalign{\kern-\ex@\nointerlineskip}$\m@th\hfil#1#2\hfil$\crcr}}}%
\def\underrightarrow{\mathpalette\underrightarrow@}%
\def\underrightarrow@#1#2{\vtop{\ialign{##\crcr$\m@th\hfil#1#2\hfil
  $\crcr\noalign{\nointerlineskip}\rightarrowfill@#1\crcr}}}%

\def\underleftarrow{\mathpalette\underleftarrow@}%
\def\underleftarrow@#1#2{\vtop{\ialign{##\crcr$\m@th\hfil#1#2\hfil
  $\crcr\noalign{\nointerlineskip}\leftarrowfill@#1\crcr}}}%
\def\underleftrightarrow{\mathpalette\underleftrightarrow@}%
\def\underleftrightarrow@#1#2{\vtop{\ialign{##\crcr$\m@th
  \hfil#1#2\hfil$\crcr
 \noalign{\nointerlineskip}\leftrightarrowfill@#1\crcr}}}%

\def\qopnamewl@#1{\mathop{\operator@font#1}\nlimits@}
\let\nlimits@\displaylimits
\def\setboxz@h{\setbox\z@\hbox}

\def\varlim@#1#2{\mathop{\vtop{\ialign{##\crcr
 \hfil$#1\m@th\operator@font lim$\hfil\crcr
 \noalign{\nointerlineskip}#2#1\crcr
 \noalign{\nointerlineskip\kern-\ex@}\crcr}}}}

 \def\rightarrowfill@#1{\m@th\setboxz@h{$#1-$}\ht\z@\z@
  $#1\copy\z@\mkern-6mu\cleaders
  \hbox{$#1\mkern-2mu\box\z@\mkern-2mu$}\hfill
  \mkern-6mu\mathord\rightarrow$}
\def\leftarrowfill@#1{\m@th\setboxz@h{$#1-$}\ht\z@\z@
  $#1\mathord\leftarrow\mkern-6mu\cleaders
  \hbox{$#1\mkern-2mu\copy\z@\mkern-2mu$}\hfill
  \mkern-6mu\box\z@$}

\def\projlim{\qopnamewl@{proj\,lim}}
\def\injlim{\qopnamewl@{inj\,lim}}
\def\varinjlim{\mathpalette\varlim@\rightarrowfill@}
\def\varprojlim{\mathpalette\varlim@\leftarrowfill@}
\def\varliminf{\mathpalette\varliminf@{}}
\def\varliminf@#1{\mathop{\underline{\vrule\@depth.2\ex@\@width\z@
   \hbox{$#1\m@th\operator@font lim$}}}}
\def\varlimsup{\mathpalette\varlimsup@{}}
\def\varlimsup@#1{\mathop{\overline
  {\hbox{$#1\m@th\operator@font lim$}}}}

\def\stackunder#1#2{\mathrel{\mathop{#2}\limits_{#1}}}%
\begingroup \catcode `|=0 \catcode `[= 1
\catcode`]=2 \catcode `\{=12 \catcode `\}=12
\catcode`\\=12 
|gdef|@alignverbatim#1\end{align}[#1|end[align]]
|gdef|@salignverbatim#1\end{align*}[#1|end[align*]]

|gdef|@alignatverbatim#1\end{alignat}[#1|end[alignat]]
|gdef|@salignatverbatim#1\end{alignat*}[#1|end[alignat*]]

|gdef|@xalignatverbatim#1\end{xalignat}[#1|end[xalignat]]
|gdef|@sxalignatverbatim#1\end{xalignat*}[#1|end[xalignat*]]

|gdef|@gatherverbatim#1\end{gather}[#1|end[gather]]
|gdef|@sgatherverbatim#1\end{gather*}[#1|end[gather*]]

|gdef|@gatherverbatim#1\end{gather}[#1|end[gather]]
|gdef|@sgatherverbatim#1\end{gather*}[#1|end[gather*]]

|gdef|@multilineverbatim#1\end{multiline}[#1|end[multiline]]
|gdef|@smultilineverbatim#1\end{multiline*}[#1|end[multiline*]]

|gdef|@arraxverbatim#1\end{arrax}[#1|end[arrax]]
|gdef|@sarraxverbatim#1\end{arrax*}[#1|end[arrax*]]

|gdef|@tabulaxverbatim#1\end{tabulax}[#1|end[tabulax]]
|gdef|@stabulaxverbatim#1\end{tabulax*}[#1|end[tabulax*]]

|endgroup

\def\align{\@verbatim \frenchspacing\@vobeyspaces \@alignverbatim
You are using the "align" environment in a style in which it is not defined.}

\@namedef{align*}{\@verbatim\@salignverbatim
You are using the "align*" environment in a style in which it is not defined.}
\expandafter\let\csname endalign*\endcsname =\endtrivlist

\def\alignat{\@verbatim \frenchspacing\@vobeyspaces \@alignatverbatim
You are using the "alignat" environment in a style in which it is not defined.}

\@namedef{alignat*}{\@verbatim\@salignatverbatim
You are using the "alignat*" environment in a style in which it is not defined.}
\expandafter\let\csname endalignat*\endcsname =\endtrivlist

\def\xalignat{\@verbatim \frenchspacing\@vobeyspaces \@xalignatverbatim
You are using the "xalignat" environment in a style in which it is not defined.}

\@namedef{xalignat*}{\@verbatim\@sxalignatverbatim
You are using the "xalignat*" environment in a style in which it is not defined.}
\expandafter\let\csname endxalignat*\endcsname =\endtrivlist

\def\gather{\@verbatim \frenchspacing\@vobeyspaces \@gatherverbatim
You are using the "gather" environment in a style in which it is not defined.}

\@namedef{gather*}{\@verbatim\@sgatherverbatim
You are using the "gather*" environment in a style in which it is not defined.}
\expandafter\let\csname endgather*\endcsname =\endtrivlist

\def\multiline{\@verbatim \frenchspacing\@vobeyspaces \@multilineverbatim
You are using the "multiline" environment in a style in which it is not defined.}

\@namedef{multiline*}{\@verbatim\@smultilineverbatim
You are using the "multiline*" environment in a style in which it is not defined.}
\expandafter\let\csname endmultiline*\endcsname =\endtrivlist

\def\arrax{\@verbatim \frenchspacing\@vobeyspaces \@arraxverbatim
You are using a type of "array" construct that is only allowed in AmS-LaTeX.}

\def\tabulax{\@verbatim \frenchspacing\@vobeyspaces \@tabulaxverbatim
You are using a type of "tabular" construct that is only allowed in AmS-LaTeX.}

\@namedef{arrax*}{\@verbatim\@sarraxverbatim
You are using a type of "array*" construct that is only allowed in AmS-LaTeX.}
\expandafter\let\csname endarrax*\endcsname =\endtrivlist

\@namedef{tabulax*}{\@verbatim\@stabulaxverbatim
You are using a type of "tabular*" construct that is only allowed in AmS-LaTeX.}
\expandafter\let\csname endtabulax*\endcsname =\endtrivlist

\def\@@eqncr{\let\@tempa\relax
    \ifcase\@eqcnt \def\@tempa{& & &}\or \def\@tempa{& &}%
      \else \def\@tempa{&}\fi
     \@tempa
     \if@eqnsw
        \iftag@
           \@taggnum
        \else
           \@eqnnum\stepcounter{equation}%
        \fi
     \fi
     \global\tag@false
     \global\@eqnswtrue
     \global\@eqcnt\z@\cr}

 \def\endequation{%
     \ifmmode\ifinner 
      \iftag@
        \addtocounter{equation}{-1} 
        $\hfil
           \displaywidth\linewidth\@taggnum\egroup \endtrivlist
        \global\tag@false
        \global\@ignoretrue   
      \else
        $\hfil
           \displaywidth\linewidth\@eqnnum\egroup \endtrivlist
        \global\tag@false
        \global\@ignoretrue 
      \fi
     \else   
      \iftag@
        \addtocounter{equation}{-1} 
        \eqno \hbox{\@taggnum}
        \global\tag@false%
        $$\global\@ignoretrue
      \else
        \eqno \hbox{\@eqnnum}
        $$\global\@ignoretrue
      \fi
     \fi\fi
 } 

 \newif\iftag@ \tag@false
 
 \def\tag{\@ifnextchar*{\@tagstar}{\@tag}}
 \def\@tag#1{%
     \global\tag@true
     \global\def\@taggnum{(#1)}}
 \def\@tagstar*#1{%
     \global\tag@true
     \global\def\@taggnum{#1}%
}

\makeatother

\begin{document}

\begin{center}
\textbf{CORRELATION BETWEEN THE DEUTERON CHARACTERISTICS AND THE LOW-ENERGY
TRIPLET }$\mathbf{np}$ \textbf{SCATTERING} \textbf{PARAMETERS}

\textbf{\\[0pt]
V. A. Babenko \footnote[1]{{\normalsize
\mbox{Electronic address:
pet@gluk.org}}}
 and N. M. Petrov}\\[0pt]
\textit{Bogolyubov Institute for Theoretical Physics,}

\textit{National Academy of Sciences of Ukraine, Kiev, Ukraine}
\end{center}

\vspace{1pt}

\noindent The correlation relationship between the deuteron asymptotic
normalization constant, $A_{S}$, and the triplet $np$ scattering length, $%
a_{t}$, is investigated. It is found that $99.7\%$ of the asymptotic
constant $A_{S}$ is determined by the scattering length $a_{t}$. It is shown
that the linear correlation relationship between the quantities $A_{S}^{-2}$
and $1/a_{t}$ provides a good test of correctness of various models of
nucleon-nucleon interaction. It is revealed that, for the normalization
constant $A_{S}$ and for the root-mean-square deuteron radius $r_{d}$, the
results obtained with the experimental value recommended at present for the
triplet scattering length $a_{t}$ are exaggerated with respect to their
experimental counterparts. By using the latest experimental phase shifts of
Arndt et al., we obtain, for the low-energy scattering parameters ($a_{t}$, $%
r_{t}$, $P_{t}$) and for the deuteron characteristics ($A_{S}$, $r_{d}$),
results that comply well with experimental data.

\vspace{1pt}\noindent

\vspace{12pt}

\textbf{1.} Basic features of the deuteron --- such as the binding energy $%
\varepsilon _{d}$; the electric quadrupole moment $Q$; the root-mean-square
radius $r_{d}$; the asymptotic normalization constants for the $S$ and the $D
$ wave, $A_{S}$ and $A_{D}$; and the corresponding asymptotic $D/S$ ratio $%
\eta =A_{D}/A_{S}$ ---\ play a significant role in constructing realistic
models of nucleon-nucleon interaction and are important physical
characteristics of nuclear forces. Of equally great importance are
low-energy $np$ scattering parameters in the triplet state. This include the
scattering length $a_{t}$; the effective range $r_{t}$; and the shape
parameters $v_{2}$, $v_{3}$, $v_{4}$, ... appearing in the effective-range
expansion
\begin{equation}
k\cot \delta _{t}\left( k\right) =-\frac{1}{a_{t}}+\frac{1}{2}%
r_{t}k^{2}+v_{2}k^{4}+v_{3}k^{6}+v_{4}k^{8}+\ldots \,,  \tag{1}
\end{equation}
where $\delta _{t}\left( k\right) $ is the triplet $np$ scattering
eigenphase corresponding to the $^{3}S_{1}$ state. For this reason, much
attention has been given to these quantities both in theoretical and in
experimental studies \lbrack 1--17\rbrack . At the present time, the
experimental value of the deuteron binding energy $\varepsilon _{d}$ is
known to a high precision \lbrack 9\rbrack :
\begin{equation}
\varepsilon _{d}=2.22458900\,MeV\,.  \tag{2}
\end{equation}
The value of the asymptotic $D/S$ ratio, $\eta $, was also determined to a
fairly good precision, both theoretically and experimentally (see \lbrack 4,
7, 15--17\rbrack ). The majority of the theoretical estimates of this
quantity are in good agreement with its experimental value of $\eta =0.0272\,
$ \lbrack 15\rbrack . At the same time, the values of some characteristics
of the deuteron, such as the asymptotic normalization constant $A_{S}$ and
the root-mean-square radius $r_{d}$, were the subject of controversy. For
example, the value obtained for $A_{S}$ directly from experimental data by
analyzing elastic $pd$ scattering \lbrack 18\rbrack ,
\begin{equation}
A_{S}=0.8781\,fm^{-1/2}\,\,,  \tag{3}
\end{equation}
is at odds with the theoretical estimates derived for this quantity for many
realistic potentials \lbrack 19--26\rbrack , as well as with those found
from an analysis of phase shifts \lbrack 17, 27\rbrack\ and on the basis of
the effective-range expansion \lbrack 4, 8\rbrack . Values of the asymptotic
constant $A_{S}$ that are discussed in the literature vary within a rather
broad range --- from $0.7592\,\,fm^{-1/2}$ to $0.9863\,\,fm^{-1/2}$ \lbrack
28, 29\rbrack .

\vspace{1pt}

In a number of studies the result reported by Ericson in \lbrack 5\rbrack , $%
A_{S}=0.8802\,fm^{-1/2}$, is used for an ''experimental'' value. It should
be noted, however, that this value was obtained from an analysis of the
linear relationship between the asymptotic constant $A_{S}$ and the
root-mean-square radius $r_{d}$ rather than from experimental data directly.
This linear correlation relationship between $A_{S}$ and $r_{d}$ was
established empirically for various models of nucleon-nucleon interaction
\lbrack 5, 30\rbrack . The procedure that Ericson used to obtain the above
value of $A_{S}=0.8802\,fm^{-1/2}$ on the basis of this relationship
involved averaging the values of the root-mean-square radius $r_{d}$ that
were found experimentally in \lbrack 10, 11\rbrack . In view of this, the
use of Ericson's result for an experimental value is not quite correct. The
value that is presented in (3) and which was derived in \lbrack 18\rbrack\
on the basis of a direct method for determining this normalization constant
from an analysis of data on elastic $pd$ scattering is more justified.

\vspace{1pt}

\vspace{1pt}Other values sometimes used for the experimental asymptotic
normalization constant include $A_{S}=0.8846$, $0.8848$, and $0.8883$%
\thinspace $fm^{-1/2}$ (see \lbrack 4\rbrack , \lbrack 30, 31\rbrack , and
\lbrack 8\rbrack , respectively). In just the same way as Ericson's result,
they can hardly be treated, however, as correct experimental values, since
they were found in the effective-range approximation with allowance for some
corrections associated with the form of interaction; moreover, the
low-energy triplet $np$ scattering parameters that were employed in doing
this are not determined from experimental data unambiguously. By way of
example, we indicate that, in \lbrack 1--3, 12--14, 32\rbrack , values
between $5.377\,fm$ \lbrack 1\rbrack\ and $5.424\,fm$ \lbrack 32\rbrack\ are
given for experimental values of the triplet scattering length $a_{t}$, but
the asymptotic normalization constant $A_{S}$ greatly depends on $a_{t}$. As
will be shown below, $99.7\%$ of the normalization constant $A_{S}$ is
determined by the triplet scattering length $a_{t}$.

\vspace{1pt}

\vspace{1pt}In the following, the value in (3) from \lbrack 18\rbrack\ will
be used for the experimental value of the normalization constant $A_{S}$. It
is close to the value of $A_{S}=0.8771\,fm^{-1/2}$, which corresponds to the
vertex-constant value of $G_{d}^{2}=0.427\,fm$ for the $d\rightarrow n+p$
vertex and which was obtained earlier in \lbrack 33\rbrack . As we have
already indicated, the authors of \lbrack 18, 33\rbrack\ employed a direct
method for determining the constants $A_{S}$ and $G_{d}^{2}$ that relies on
extrapolating the experimental cross sections for elastic $pd$ scattering to
the exchange-singularity point.

\vspace{1pt}

In just the same way as the constant $A_{S}$, the root-mean-square radius $%
r_{d}$ of the deuteron is determined, to great extent, by the triplet
scattering length $a_{t}$. A linear correlation relationship between the
quantities $r_{d}$ and $a_{t}$ was established empirically in \lbrack 6,
30\rbrack . To a high precision, this relationship can be approximated as
follows:
\begin{equation}
r_{d}=0.4a_{t}-0.1985\,\,\,\left( fm\right) \,.  \tag{4}
\end{equation}
At present, the following experimental values are used in the literature for
the root-mean-square radius of the deuteron:
\begin{equation}
r_{d}=1.9635\,fm\text{\ \lbrack 10\rbrack\ \ \ \ (B),}  \tag{5a}
\end{equation}
\begin{equation}
r_{d}=1.9560\,fm\text{\ \lbrack 11\rbrack\ \ \ \ (S),}  \tag{5b}
\end{equation}
\begin{equation}
r_{d}=1.950\,fm\text{\ \lbrack 30\rbrack\ \ \ \ (K).}  \tag{5c}
\end{equation}
According to Eq. (4), the following values of the scattering length $a_{t}$
correspond to the values of the radius $r_{d}$ in (5):
\begin{equation}
a_{t}=5.4050\,fm\ \ \ \ \text{(B),}  \tag{6a}
\end{equation}
\begin{equation}
a_{t}=5.3863\,fm\ \ \ \ \text{(S),}  \tag{6b}
\end{equation}
\begin{equation}
a_{t}=5.3713\,fm\ \ \ \ \text{(K).}  \tag{6c}
\end{equation}
On the other hand, the values of the triplet scattering length that were
calculated for many realistic nucleon-nucleon potentials \lbrack 19--21, 24,
25\rbrack\ are close to the experimental value \lbrack 32\rbrack
\begin{equation}
a_{t}=5.424\,fm\,,  \tag{7}
\end{equation}
which is recommended at present, but which, in accordance with Eq. (4),
leads to the root-mean-square radius exaggerated in relation to the
experimental values in (5a)--(5c); that is,
\begin{equation}
r_{d}=1.9711\,fm\,.  \tag{8}
\end{equation}
Thus, we can see that, frequently, values obtained and used in various
studies for the characteristics of the deuteron and for the triplet
low-energy $np$ scattering parameters are contradictory and deviate from
experimental results.

\textbf{2.} In accordance with \lbrack 8\rbrack , the asymptotic
normalization constant $A_{S}$ for the deuteron can be represented in the
form
\begin{equation}
A_{S}^{2}=\frac{2\alpha }{1-\alpha \rho _{d}}\,,  \tag{9}
\end{equation}
where $\alpha $ is the deuteron wave number defined according to the
relation $\varepsilon _{d}=\hbar ^{2}\alpha ^{2}/m_{N\text{ }}$ and $\rho
_{d}\equiv \rho \left( -\varepsilon _{d},-\varepsilon _{d}\right) $ is the
deuteron effective range corresponding to $S$-wave interaction. The
definition and properties of the radius $\rho _{d}$ and of the function $%
\rho \left( E_{1},E_{2}\right) $ are discussed in detail elsewhere \lbrack
1\rbrack . The quantity $\rho _{d}$ appears in the expansion of the function
$k\cot \delta _{t}\left( k\right) $ at the point $k^{2}=-\alpha ^{2}$ ---
that is, at the energy value equal to the deuteron binding energy. The
expansion at the point $k^{2}=-\alpha ^{2}$ is similar to the expansion in
(1), which is performed at the origin, involving the ordinary effective
range $r_{t}\equiv \rho \left( 0,0\right) $ of scattering theory --- that
is, the effective range at zero energy. It can easily be found that the
quantities $\rho _{d}$ and $r_{t}$ are expanded in powers of the parameter $%
\alpha ^{2}$ as
\begin{equation}
\rho _{d}=\rho _{m}-2v_{2}\alpha ^{2}+4v_{3}\alpha ^{4}-6v_{4}\alpha
^{6}+\ldots \,,  \tag{10}
\end{equation}
\begin{equation}
r_{t}=\rho _{m}+2v_{2}\alpha ^{2}-2v_{3}\alpha ^{4}+2v_{4}\alpha ^{6}-\ldots
\,,  \tag{11}
\end{equation}
where $\rho _{m}\equiv \rho \left( 0,-\varepsilon _{d}\right) $ is the
so-called mixed effective range \lbrack 1\rbrack , for which the following
relation holds:
\begin{equation}
\rho _{m}=\frac{2}{\alpha }\left( 1-\frac{1}{\alpha a_{t}}\right) \,.
\tag{12}
\end{equation}
The shape parameters $v_{n}$ in expansions (1), (10), and (11) are
dimensional quantities. Instead of them, one \ \ \ \ \ often introduces the
dimensionless shape parameters $P_{t}$, $Q_{t}$, ... related to the
parameters $v_{n}$ by the equations
\begin{equation}
v_{2}=-P_{t}r_{t}^{3}\,,\,\,\,\,v_{3}=Q_{t}r_{t}^{5}\,,\ldots \,\,\,.
\tag{13}
\end{equation}
From the expansions in (10) and (11), it follows that the quantities $r_{t}$%
, $\rho _{d}$, and $\rho _{m}$ are related as
\begin{equation}
\rho _{d}+r_{t}=2\rho _{m}+2v_{3}\alpha ^{4}-4v_{4}\alpha ^{6}+\ldots \,,
\tag{14}
\end{equation}
\begin{equation}
\rho _{d}+2r_{t}=3\rho _{m}+2v_{2}\alpha ^{2}-2v_{4}\alpha ^{6}+\ldots \,.
\tag{15}
\end{equation}

\vspace{1pt}

The asymptotic normalization constant $A_{S}$ for the deuteron is directly
expressed in terms of the residue of the $S$ matrix $S\left( k\right) $ at
the pole \ $k=i\alpha $ corresponding to a bound state of the two-nucleon
system; that is,
\begin{equation}
A_{S}^{2}=i\stackunder{k=i\alpha }{\limfunc{Res}}S\left( k\right) \,.
\tag{16}
\end{equation}
Along with the constant $A_{S}$, other physical quantities, such as the
nuclear vertex constant $G_{d}$ and the dimensionless asymptotic
normalization constant $C_{d}$, are frequently used in the literature
\lbrack 28, 29\rbrack . These two quantities are directly related to the
constant $A_{S}$ by the equations
\begin{equation}
G_{d}^{2}=\pi \rule[5pt]{4.6pt}{0.5pt}\hspace{-0.5em}\lambda ^{2}A_{S}^{2}\,,
\tag{17}
\end{equation}
\begin{equation}
C_{d}^{2}=A_{S}^{2}/2\alpha \,,  \tag{18}
\end{equation}
where $\rule[5pt]{4.6pt}{0.5pt}\hspace{-0.5em}\lambda =2%
\rule[5pt]{4.6pt}{0.5pt}\hspace{-0.5em}\lambda _{N}\,$, with $%
\rule[5pt]{4.6pt}{0.5pt}\hspace{-0.5em}\lambda _{N}\equiv \hbar /m_{N}c$
being the Compton wavelength of the nucleon.

\vspace{1pt}

In the approximation where there is no dependence on the form of interaction
($v_{2}=v_{3}=\ldots =0$),
\begin{equation}
k\cot \delta _{t}\left( k\right) =-\frac{1}{a_{t}}+\frac{1}{2}r_{t}k^{2}\,,
\tag{19}
\end{equation}
the quantities $r_{t}$, $\rho _{d}$, and $\rho _{m}$ are equal to each
other, as follows from Eqs. (10) and (11):
\begin{equation}
r_{t}=\rho _{d}=\rho _{m}\,.  \tag{20}
\end{equation}
\vspace{1pt}In this approximation, the use of relations (12) and (20) makes
it possible to recast Eq. (9) into the more convenient form
\begin{equation}
C_{d}^{-2}=-1+2\frac{R_{d}}{a_{t}}\,,  \tag{21}
\end{equation}
where the quantity
\begin{equation}
R_{d}\equiv \frac{1}{\alpha }\,,  \tag{22}
\end{equation}
which characterizes the spatial dimensions of the deuteron, is referred to
as the deuteron radius \lbrack 2\rbrack . By using the value in (2) for the
deuteron binding energy, one can easily obtain the following numerical value
for the deuteron radius:
\begin{equation}
R_{d}=4.317688\,fm\,.  \tag{23}
\end{equation}

\vspace{1pt}

In the effective-range approximation (19) the low-energy triplet $np$
scattering parameters $a_{t}$ and $r_{t}$ are expressed in terms of the
bound-state parameters (deuteron radius $R_{d}$ and normalization constant $%
C_{d}$) as
\begin{equation}
a_{t}=\frac{2R_{d}}{1+C_{d}^{-2}}\,,  \tag{24}
\end{equation}
\begin{equation}
r_{t}=R_{d}\left( 1-C_{d}^{-2}\right) \,.  \tag{25}
\end{equation}
With the aid of the deuteron parameters, the behavior of the phase shift at
low energies can be predicted in the effective-range approximation (19) with
allowance for expressions (24) and (25).

\vspace{1pt}

\textbf{3.} As was indicated above, a great number of studies have been
devoted to exploring and calculating the asymptotic normalization constant $%
A_{S}$. The values of the constant $A_{S}$ (and of the quantities $%
\varepsilon _{d}$ and $a_{t}$) for some realistic potentials \lbrack 19--25,
27, 34--36\rbrack\ are quoted in Table 1, along with the values of $A_{S}$
that were found from the analysis of phase shifts in \lbrack 17, 27\rbrack\
and on the basis of the effective-range expansion in \lbrack 4, 8\rbrack .
Also given in the same table are the values of the constant $A_{S}$ that
were calculated in the present study by formulas (18) and (21), which
correspond to the effective-range approximation. In addition, Table 1
presents the absolute ($\Delta $) and the relative ($\delta $) error in the
calculation of $A_{S}$ in this approximation. \ \ \ \ \ \ \

\vspace{1pt}

It can be seen from Table 1 that, for the majority of the models, the
relative error in the $A_{S}$ value calculated in the effective-range
approximation does not exceed 0.3\%, while the absolute error is not greater
than $0.003\,fm^{-1/2}$, as a rule. This is not so only for some early
models of the Bonn potential (lines 15--17 in Table 1), in which case the
relative error in the approximate value of the constant $A_{S}$ is 2 to 3\%.
For the same potentials, the relative error in the approximate values of the
deuteron mean-square radius $r_{d}$ that were obtained by formula (4) is
also overly large. For example, the relative error in $r_{d}$ is 2.52\% for
the HM-2 potential and 1.49\% for the Bonn F potential. At the same time,
this error is small for more correct models of the Bonn potential, Bonn R
and Bonn Q ($0.081$ and $0.137\%$, respectively). For the Paris potential,
the error in question is $0.035\%$. \ \ \ \ \ \

\vspace{1pt}

Thus, we can see from Table 1 that the values of the asymptotic
normalization constant $A_{S}$ for realistic potentials are strongly
correlated with the values of the triplet $np$ scattering length $a_{t}$.
The same is also true for the values of $A_{S}$ that were found from the
analysis of phase shifts (lines 18, 19) and on the basis of the
effective-range expansion in \lbrack 4\rbrack\ (line 20). As to the value of
$A_{S}=0.8883\,fm^{-1/2}$ (line 21), which was calculated in \lbrack
8\rbrack\ on the basis of the effective-range expansion, it is strongly
overestimated in relation to the value of $A_{S}=0.88191\,fm^{-1/2}$, which
was found in the present study in the effective-range approximation. In that
case, the relative error in the approximate value was $0.725\%$. An
incorrect choice of the shape parameter $P_{t}$ in \lbrack 8\rbrack\ is the
reason for this discrepancy --- namely, the following data from \lbrack
37\rbrack\ on the low-energy scattering parameters were used in \lbrack
8\rbrack\ to calculate the constant $A_{S}$: $a_{t}=5.412\,fm$, and $%
r_{t}=1.733\,fm$; however, the value of $P_{t}=-0.0188$, chosen there for
the shape parameter, corresponded to the Paris potential, for which the
scattering length and the effective range take values ($a_{t}=5.427\,fm$, $%
r_{t}=1.766\,fm$) that exceed considerably those that were employed in
\lbrack 8\rbrack . \ \ \

\vspace{1pt}

Considering that the deuteron binding energy has been determined to a high
degree of precision and is taken to have an approximately the same value in
all of the calculations, one can conclude on the basis of the results in
Table 1 that $99.7\%$ of the asymptotic normalization constant $A_{S}$ is
determined by the triplet scattering length $a_{t}$. The inverse also holds:
knowing the values of the constant $A_{S}$, one can determine the triplet
scattering length $a_{t}$ to a high degree of precision.

\vspace{1pt}

Taking the aforesaid into consideration, we will investigate the asymptotic
constant $A_{S}$ as a function of the triplet scattering length $a_{t}$. It
is convenient to perform this investigation for the dimensionless quantity $%
C_{d}^{-2}$, which, in the effective-range approximation, is a linear
function of the dimensionless quantity $R_{d}/a_{t}$ \lbrack see Eq.
(21)\rbrack . The $R_{d}/a_{t}$ dependence of $C_{d}^{-2}$ is shown in the
figure. The straight line represents the results of the calculation based on
the approximate formula (21), while the points in the figure correspond to $%
\varepsilon _{d}$, $a_{t}$, and $A_{S}$ values computed by various authors
and quoted in Table 1. As can be seen from the figure, there is a linear
relationship between the quantities $C_{d}^{-2}$ and $R_{d}/a_{t}$. Points
corresponding to their values lie in the close proximity of the straight
line specified by Eq. (21). As was mentioned above, this is not so only for
points corresponding to early models of the Bonn potential and the point
corresponding to the values of the quantities in question from \lbrack
8\rbrack . Points that represent values of $C_{d}^{-2}$ and $R_{d}/a_{t}$
for more correct versions of the Bonn potential model (Bonn R, Bonn Q) lie
near the straight line specified by Eq. (21) (points 13, 14).

\vspace{1pt}

Thus, it follows from Table 1 and from the figure that the asymptotic
normalization constant $A_{S}$ and the triplet scattering length $a_{t}$ are
well correlated quantities, so that any of these can be determined to a high
degree of precision if the other is known. For the experimental value
presented in (3) for the constant $A_{S}$, the corresponding value of the
scattering length $a_{t}$ can easily be determined in the effective-range
approximation by formulas (18) and (21). The result is
\begin{equation}
a_{t}=5.395\,fm\,.  \tag{26}
\end{equation}
At the same time, the currently recommended experimental value of the
triplet scattering length in (7) leads, in the effective-range
approximation, to the asymptotic-normalization-constant value
\begin{equation}
A_{S}=0.88451\,fm^{-1/2}\,\,,  \tag{27}
\end{equation}
which is well above the experimental value of this quantity in (3). As was
indicated above, the experimental scattering-length value in (7) also leads
to the exaggerated value in (8) for the root-mean-square radius $r_{d}$ of
the deuteron.

\vspace{1pt}

Thus, it can be concluded from the above analysis that the currently
recommended experimental value of the triplet $np$ scattering length in (7)
does not comply with the experimental values of the asymptotic normalization
constant $A_{S}$ for the deuteron and its root-mean-square radius $r_{d}$ in
(3) and (5a)--(5c), respectively. Therefore, it is of paramount importance
to determine, for the characteristics of the deuteron and for the low-energy
triplet $np$ scattering parameters, such values that would be consistent
with one another, on one hand, and which would be compatible with
experimental data on the other hand.

\vspace{1pt}

\textbf{4.} Fixing the features $\varepsilon _{d}$, $A_{S}$, and $r_{d}$ of
the deuteron, we will study the behavior of the $S$-wave phase shift at low
energies in the approximation that takes into account the shape parameter $%
P_{t}$ in the effective-range expansion; that is,
\begin{equation}
k\cot \delta _{t}\left( k\right) =-\frac{1}{a_{t}}+\frac{1}{2}%
r_{t}k^{2}-P_{t}r_{t}^{3}k^{4}\,.  \tag{28}
\end{equation}
For the scattering length $a_{t}$, use is made here of the values in
(6a)--(6c), which correspond to the values of the root-mean-square-radius $%
r_{d}$ of the deuteron in (5a)--(5c), while, in accordance with (10), (13),
and (14), the effective range $r_{t}$ and the shape parameter $P_{t}$ are
given by
\begin{equation}
r_{t}=2\rho _{m}-\rho _{d}\,,  \tag{29}
\end{equation}
\begin{equation}
P_{t}=\frac{\rho _{d}-\rho _{m}}{2r_{t}^{3}\alpha ^{2}}\,,  \tag{30}
\end{equation}
where the effective radius $\rho _{d}$ of the deuteron and the mixed
effective range $\rho _{m}$ are determined from Eqs. (9) and (12),
respectively. The parameters $a_{t}$, $r_{t}$,\ $P_{t}$, $\rho _{m}$, and $%
\rho _{d}$ calculated in this approximation on the basis of the experimental
values of the deuteron binding energy $\varepsilon _{d}$ in (2), the
asymptotic normalization constant $A_{S}$ in (3), and the root-mean-square
radius $r_{d}$ of the deuteron in (5a)--(5c) are given in Table 2 (B, S, K),
along with the values calculated for these parameters in the effective-range
approximation (ER) on the basis of the experimental values of $\varepsilon
_{d}$ and $A_{S}$, as well as values found in the present study from the
latest partial-wave analysis (PWA) of the $NN$ scattering by Arndt et al.
\lbrack 38\rbrack , which, as can be seen from this table, are in very good
agreement with the corresponding values for version B, where the features of
the deuteron were set to $\varepsilon _{d}=2.22458900$\thinspace $MeV$
\lbrack 9\rbrack , $A_{S}=0.8781\,fm^{-1/2}$ \lbrack 18\rbrack , and $%
r_{d}=1.9635\,fm$ \lbrack 10\rbrack .

\vspace{1pt}

The quantity $\delta \rho $ defined as the difference of the effective
radius $\rho _{d}$ of the deuteron and the mixed effective range $\rho _{m}$%
,
\begin{equation}
\delta \rho =\rho _{d}-\rho _{m}\,,  \tag{31}
\end{equation}
is often discussed in the literature. According to the estimates obtained by
Noyes in \lbrack 3\rbrack\ on the basis of dispersion relations, this
difference arises owing to one-pion exchange and is positive, its magnitude
being $0.016\,fm$. For many potential models, the difference $\delta \rho $
is also positive; as was established in \lbrack 5, 6, 31\rbrack , it well
correlates with the triplet scattering length $a_{t}$. In our case, this
difference is given by
\begin{equation}
\delta \rho =2P_{t}r_{t}^{3}\alpha ^{2}.  \tag{32}
\end{equation}
For the parameter values used in the S and K versions from Table 2, it is
positive, taking the values of $0.011$ and $0.030\,fm$, respectively.
However, for the cases of B and PWA in Table 2, the difference $\delta \rho $
is negative, its values being $-0.013$ and $-0.015\,fm$. For this reason,
the problem of the sign and magnitude of the difference of the effective
radius of the deuteron $\rho _{d}$ and the mixed effective range $\rho _{m}$
calls for a further investigation.

\vspace{1pt}

For the low-energy parameters given in Table 2, we have calculated the
triplet phase shift $\delta _{t}\left( k\right) $. The results are displayed
in Table 3, along with the latest experimental data of Arndt et al. \lbrack
38\rbrack\ on the triplet phase shift. Table 4 presents the energy
dependence of the difference
\begin{equation}
\Delta =\delta _{\exp }-\delta _{theor}\,  \tag{33}
\end{equation}
of the experimental value of the phase shift and its theoretical
counterparts calculated by formula (28) and quoted in Table 3. As can be
seen from Tables 3 and 4, all sets of low-energy parameters $a_{t}$, $r_{t}$%
, and $P_{t}$ from Table 2 describe well experimental data up to an energy
value of $5\,MeV$ (the absolute error being less than $1^{\circ }$).
Nonetheless, the distinction between the sets of low-energy parameters in
describing experimental data becomes noticeable at an energy as low as $%
1\,MeV$, and we can see that preference should be given to the parameter
sets employed in the B and PWA versions. These sets provide a nearly
precision description (with a relative error of about $0.005\%$) of
experimental phase shifts up to an energy value of $5\,MeV$. Thus, the
accuracy of existing experimental data that is available at present is quite
sufficient for removing ambiguities that arise in determining the scattering
length $a_{t}$ and the effective range $r_{t}$ \lbrack 1\rbrack\ and which
are associated with the form of the potential. In view of this, the problem
of deducing the scattering length, the effective range, and parameters of
higher order (shape parameters) in expansion (1) directly from experimental
data is pressing. It should be noted that the B and PWA sets describe well,
in contrast to other parameter sets, experimental phase shifts up to an
energy value of $50\,MeV$. For the B and PWA sets, the absolute error is
about $0.5^{\circ }$ at $E_{lab}=30\,MeV$ and about $1^{\circ }$ at $%
E_{lab}=50\,MeV$. From Table 4, it can be seen that, for other parameter
sets, the absolute error is much greater.

\vspace{1pt}

Thus, $np$ scattering in the triplet state can be described rather well
within the B set up to an energy value of $50\,MeV$, the experimental values
used in this description for the characteristics of the deuteron being that
in (2) for the binding energy, that in (3) for the asymptotic normalization
constant, and that in (5a) for the root-mean-square radius. On the other
hand, parameters that characterize the neutron-proton bound state (deuteron)
can be determined from experimental data on $np$ scattering. By using the
values
\begin{equation}
a_{t}=5.4030\,fm\,,\text{ }r_{t}=1.7495\,fm\,,\text{ }P_{t}=-0.0259\,,
\tag{34}
\end{equation}
which we found here for the low-energy scattering parameters from an
analysis of the latest data on phase shifts \lbrack 38\rbrack , we will now
determine the asymptotic normalization constant $A_{S}$ and the
root-mean-square radius $r_{d}$ for the deuteron. In accordance with Eqs.
(4), (9), (10), (12), and (13), we obtain
\begin{equation}
A_{S}=0.8774\,fm^{-1/2}\,,  \tag{35}
\end{equation}
\begin{equation}
\,r_{d}=1.9627\,fm\,.  \tag{36}
\end{equation}
As might have been expected, these results are in very good agreement with
the experimental values of $A_{S}=0.8781\,fm^{-1/2}$\ \lbrack 18\rbrack\ and
$r_{d}=1.9635$\thinspace $fm$ \lbrack 10\rbrack .

\vspace{1pt}

\textbf{5.} To summarize, we will formulate our basic results and
conclusions. We have investigated the correlation relationship between the
asymptotic normalization constant for the deuteron, $A_{S}$, and the triplet
\ $np$ scattering length $a_{t}$. It has been established that $99.7\%$ of
the asymptotic constant $A_{S}$ is determined by the triplet scattering
length $a_{t}$. It has been shown that, in the effective-range
approximation, the linear correlation relationship between the quantities $%
2\alpha /A_{S}^{2}$ and $R_{d}/a_{t}$ provides a good test of correctness of
various potential models and methods that are used in studying
nucleon-nucleon interaction.

\vspace{1pt}

It has been found that evaluating the asymptotic normalization constant for
the deuteron and its root-mean-square radius with the currently recommended
triplet-scattering-length value of $a_{t}=5.424\,fm$ \lbrack 32\rbrack\
leads to results, $A_{S}\simeq 0.8845\,fm^{-1/2}$ and $r_{d}=1.9711\,fm$,
that are exaggerated in relation to the corresponding experimental values in
(3) and (5a)--(5c).

\vspace{1pt}

By using the experimental values of $\varepsilon _{d}=2.22458900\,MeV$
\lbrack 9\rbrack , $A_{S}=0.8781\,fm^{-1/2}$ \lbrack 18\rbrack , and $%
r_{d}=1.9635\,fm$ \lbrack 1\rbrack\ for, respectively, the binding energy of
the deuteron, its asymptotic normalization constant, and its
root-mean-square radius, we have obtained the following results for the
low-energy scattering parameters: $a_{t}=5.4050\,fm\,$, $r_{t}=1.7505\,fm$,
and $P_{t}=-0.0231$. It turned out that, with these parameter values, the
respective experimental phase shift is faithfully reproduced up to an energy
value of $50\,MeV$. The absolute error in the phase shift at this energy
value is about $1^{\circ }$. As the energy decreases, the absolute error
becomes smaller, taking the value of $0.06^{\circ }$ at an energy of $1\,MeV$%
. If the root-mean-square radius of the deuteron is set to the experimental
value of $r_{d}=1.9560\,fm$ from \lbrack 11\rbrack\ or the experimental
value of $r_{d}=1.950\,fm$ from \lbrack 3\rbrack , the corresponding
low-energy scattering parameters lead to a much poorer description of the
experimental phase shift in the shape-parameter approximation. Even at an
energy of $10\,MeV$, the absolute error exceeds $1^{\circ }$ in this case.

\vspace{1pt}

The values found for the low-energy scattering parameters with the aid of
the latest experimental results of Arndt et al. \lbrack 38\rbrack\ for the
phase shifts are $a_{t}=5.4030\,fm$, $r_{t}=1.7495\,fm$, and $P_{t}=-0.0259$%
. They are in good agreement with the parameter values of $a_{t}=5.4050\,fm$%
, $r_{t}=1.7505\,fm$, and $P_{t}=-0.0231$, which were obtained on the basis
of the experimental values of the characteristics of the deuteron. Both sets
of these parameters make it possible to describe well, in the
shape-parameter approximation, the experimental phase shift up to an energy
of $50\,MeV$, this indicating that the parameter $Q_{t}$ and parameters of
higher order in the effective-range expansion are small.

\vspace{1pt}

On the basis of the parameter values of $a_{t}=5.4030\,fm$, $%
r_{t}=1.7495\,fm $, and $P_{t}=-0.0259$, which correspond to the
experimental phase shifts obtained by Arndt et al. \lbrack 38\rbrack , we
have found the asymptotic normalization constant for the deuteron, $%
A_{S}=0.8774\,fm^{-1/2}$, and its root-mean-square radius, $r_{d}=1.9627\,fm$%
, these results being in excellent agreement with the experimental values of
$A_{S}=0.8781\,fm^{-1/2}$ \lbrack 18\rbrack\ and $r_{d}=1.9635\,fm$ \lbrack
10\rbrack . \ \ \ \ \ \ \

\vspace{1pt}

In summary, we arrive at the basic conclusion that the latest experimental
results of Arndt et al. \lbrack 38\rbrack\ for the phase shifts comply very
well with the experimental values of parameters that characterize the
deuteron, specifically, with the binding energy in (2), the asymptotic
normalization constant determined in \lbrack 18\rbrack\ and given in (3),
and the root-mean-square radius of the deuteron as obtained in \lbrack
10\rbrack\ and presented in (5a).

\vspace{1pt}\vspace{12pt}

\begin{center}
REFERENCES
\end{center}

\begin{enumerate}
\item  L. Hulth\'{e}n and M. Sugawara, in \textit{Handbuch der Physik}, Ed.
by S. Fl\"{u}gge (Springer-Verlag, New York, Berlin, 1957), p. 1.

\item  R. Wilson, \textit{The Nucleon-Nucleon Interaction} (Interscience,
New York, 1963).

\item  H. P. Noyes, Ann. Rev. Nucl. Sci. \textbf{22}, 465 (1972).

\item  T. E. O. Ericson and M. Rosa-Clot, Nucl. Phys. \textbf{A405}, 497\
(1983).

\item  T. E. O. Ericson, Nucl. Phys. \textbf{A416, }281 (1984).

\item  D. W. L. Sprung, in \textit{Proceedings of IX European Conference on
Few-Body Problems in Physics, Tbilisi, Georgia, USSR, Aug. 1984,} (World
Scientific Press, Singapore, Philadelphia, 1984), p. 234.

\item  S. Klarsfeld, J. Martorell, and D. W. L. Sprung, Nucl. Phys. \textbf{%
A352}, 113 (1981).

\item  M. W. Kermode, A. McKerrell, J. P. McTavish, and L. J. Allen, Z.
Phys. \textbf{A303, }167 (1981).

\item  G. L. Greene, E. G. Kessler Jr., R. D. Deslattes, and H. Boerner,
Phys. Rev. Lett. \textbf{56}, 819\ (1986).

\item  R. W. B\'{e}rard, F. R. Buskirk, E. B. Dally, et al., Phys. Lett.
\textbf{B47}, 355 (1973).

\item  G. G. Simon, Ch. Schmitt, and V. H. Walther, Nucl. Phys. \textbf{A364}%
, 285 (1981).

\item  T. L. Houk, Phys. Rev. \textbf{C3}, 1886\ (1971).

\item  W. Dilg, Phys. Rev. \textbf{C11}, 103\ (1975).

\item  T. L. Houk, and R. Wilson, Rev. Mod. Phys. \textbf{39}, 546\ (1967).

\item  I. Borb\'{e}ly, W. Gr\"{u}ebler, V. K\"{o}nig, et al., Phys. Lett.
\textbf{B109}, 262\ (1982).

\item  J. Hor\'{a}\v{c}ek, J. Bok, V. M. Krasnopolskij, and V. I. Kukulin,
Phys. Lett. \textbf{B172}, 1\ (1986).

\item  V. G. J. Stoks, P. C. van Campen, W. Spit, and J. J. de Swart, Phys.
Rev. Lett. \textbf{60}, 1932\ (1988).

\item  I. Borb\'{e}ly, W. Gr\"{u}ebler, V. K\"{o}nig, et al., Phys. Lett.
\textbf{B160}, 17\ (1985).

\item  M. Lacombe, B. Loiseau, J. M. Richard, et al., Phys. Rev. \textbf{C21}%
, 861\ (1980).

\item  R. Machleidt, K. Holinde, and Ch. Elster, Phys. Rep. \textbf{149}, 1\
(1987).

\item  V. G. J. Stoks, R. A. M. Klomp, C. P. F. Terheggen, and J. J. de
Swart, Phys. Rev. \textbf{C49}, 2950\ (1994).

\item  K. Holinde and R. Machleidt, Nucl. Phys. \textbf{A256}, 479\ (1976).

\item  N. J. McGurk, Phys. Rev. \textbf{C15}, 1924\ (1977).

\item  R. Machleidt, in \textit{Proceedings of IX European Conference on
Few-Body Problems in Physics, Tbilisi, Georgia, USSR, Aug. 1984,} (World
Scientific Press, Singapore, Philadelphia, 1984), p. 218.

\item  R. B. Viringa, V. G. J. Stoks, and R. Schiavilla, Phys. Rev. \textbf{%
C51}, 38\ (1995).

\item  J. W. Humberstone and J. B. G. Wallace, Nucl. Phys. \textbf{A141},
362\ (1970).

\item  J. J. de Swart, C. P. F. Terheggen, and V. G. J. Stoks, \textit{%
Invited talk at the 3rd International Symposium ''Dubna Deuteron 95'',
Dubna, Russia, July 4-7, 1995;} nucl-th/9509032.

\item  L. D. Blokhintsev, I. Borbely, and E. I. Dolinski\^{i}, Fiz. Chastits
At. Yadra \textbf{8}, 1189 (1977) \lbrack Sov. J. Part. Nucl. \textbf{8},
485 (1977)\rbrack .

\item  M. P. Locher and T. Misutani, Phys. Rep. \textbf{46}, 43\ (1978).

\item  S. Klarsfeld, J. Martorell, J. A. Oteo, M. Nishimura, D. W. L.
Sprung, Nucl. Phys. \textbf{A456}, 373\ (1986).

\item  S. Klarsfeld, J. Martorell, and\ D. W. L. Sprung, J. Phys. \textbf{G10%
}, 165\ (1984).

\item  O. Dumbrajs, R. Koch, H. Pilkuhn, et al., Nucl. Phys. \textbf{B216},
277\ (1983).

\item  A. M. Mukhamedzhanov, I. Borbely, V. Grubyuler, et al., Izv. Akad.
Nauk SSSR, Ser. Fiz. \textbf{48}, 350 (1984).

\item  R. V. Reid Jr., Ann. Phys. \textbf{50}, 411\ (1968).

\item  V. I. Kukulin, V. M. Krasnopolski\^{i}, V. N. Pomerantsev, and P. B.
Sazonov, Yad. Fiz. \textbf{43}, 559 (1986) \lbrack Sov. J. Nucl. Phys.
\textbf{43}, 355 (1986)\rbrack .

\item  N. K. Glendenning and G. Kramer, Phys. Rev. \textbf{126}, 2159 (1962).

\item  J. J. de Swart, in \textit{Few Body Problems in Nuclear and Particle
Physics}, Ed. by J. R. Slobodrian et al. (Laval, Quebec, 1975), p. 235.

\item  R. A. Arndt, W. J. Briscoe, R. L. Workman, and I. I. Strakovsky,
\textit{Partial-Wave Analysis Facility\ SAID}, URL http://gwdac.phys.gwu.edu.
\end{enumerate}

\newpage

\noindent \textbf{Table 1. }Asymptotic normalization constant $A_{S}$ for
the deuteron within various models of nucleon- nucleon interaction

\begin{center}
\vspace{14pt}

\begin{tabular}{|c|c|c|c|c|c|c|c|}
\hline
No. & References & $\varepsilon _{d}$\thinspace , & $a_{t}\,$, & $A_{S}\,$,
& $A_{S}\,$, & $\Delta \,$, & $\delta \,$, \\
&  & $MeV$ & $fm$ & $fm^{-1/2}$, & $fm^{-1/2}$, & $fm^{-1/2}$ & \% \\
&  &  &  & precise & effective-range &  &  \\
&  &  &  & value & approximation &  &  \\ \hline
1. & \multicolumn{1}{|l|}{RHC \lbrack 34\rbrack} & \multicolumn{1}{|l|}{$%
2.22464$} & $5.397$ & $0.88034$ & $0.87867$ & $0.00167$ & $0.216$ \\ \hline
2. & \multicolumn{1}{|l|}{RSC \lbrack 34\rbrack} & \multicolumn{1}{|l|}{$%
2.2246$} & $5.390$ & $0.87758$ & $0.87711$ & $0.00047$ & $0.054$ \\ \hline
3. & \multicolumn{1}{|l|}{Paris \lbrack 19\rbrack} & \multicolumn{1}{|l|}{$%
2.2249$} & $5.427$ & $0.8869$ & $0.88528$ & $0.00162$ & $0.183$ \\ \hline
4. & \multicolumn{1}{|l|}{Moscow \lbrack 35\rbrack} & \multicolumn{1}{|l|}{$%
2.2246$} & $5.413$ & $0.8814$ & $0.88211$ & $0.00071$ & $0.080$ \\ \hline
5. & \multicolumn{1}{|l|}{Nijm I\ \lbrack 21,27\rbrack} &
\multicolumn{1}{|l|}{$2.224575$} & $5.418$ & $0.8841$ & $0.88272$ & $0.00138$
& $0.156$ \\ \hline
6. & \multicolumn{1}{|l|}{Nijm II\ \lbrack 21,27\rbrack} &
\multicolumn{1}{|l|}{$2.224575$} & $5.420$ & $0.8845$ & $0.88316$ & $0.00134$
& $0.152$ \\ \hline
7. & \multicolumn{1}{|l|}{Reid 93\ \lbrack 21,27\rbrack} &
\multicolumn{1}{|l|}{$2.224575$} & $5.422$ & $0.8853$ & $0.88360$ & $0.00170$
& $0.193$ \\ \hline
8. & \multicolumn{1}{|l|}{Argonne $v_{18}$\ \lbrack 25\rbrack} &
\multicolumn{1}{|l|}{$2.224575$} & $5.419$ & $0.8850$ & $0.88341$ & $0.00159$
& $0.180$ \\ \hline
9. & \multicolumn{1}{|l|}{GK-4\ \lbrack 36\rbrack} & \multicolumn{1}{|l|}{$%
2.226$} & $5.364$ & $0.87462$ & $0.87201$ & $0.00261$ & $0.299$ \\ \hline
10. & \multicolumn{1}{|l|}{GK-8 \lbrack 36\rbrack} & \multicolumn{1}{|l|}{$%
2.226$} & $5.413$ & $0.88434$ & $0.88262$ & $0.00172$ & $0.194$ \\ \hline
11. & \multicolumn{1}{|l|}{GK-7\ \lbrack 36\rbrack} & \multicolumn{1}{|l|}{$%
2.226$} & $5.477$ & $0.89776$ & $0.89678$ & $0.00098$ & $0.109$ \\ \hline
12. & \multicolumn{1}{|l|}{HM-1 \lbrack 22,23\rbrack} & \multicolumn{1}{|l|}{%
$2.224$} & $5.50$ & $0.901$ & $0.90119$ & $0.00019$ & $0.021$ \\ \hline
13. & \multicolumn{1}{|l|}{Bonn R\ \lbrack 20\rbrack} & \multicolumn{1}{|l|}{%
$2.2246$} & $5.423$ & $0.8860$ & $0.88430$ & $0.00170$ & $0.193$ \\ \hline
14. & \multicolumn{1}{|l|}{Bonn Q\ \lbrack 20\rbrack} & \multicolumn{1}{|l|}{%
$2.2246$} & $5.424$ & $0.8862$ & $0.88452$ & $0.00168$ & $0.190$ \\ \hline
15. & \multicolumn{1}{|l|}{Bonn F\ \lbrack 20\rbrack} & \multicolumn{1}{|l|}{%
$2.2246$} & $5.427$ & $0.9046$ & $0.88517$ & $0.01942$ & $2.195$ \\ \hline
16. & \multicolumn{1}{|l|}{HM-2\ \lbrack 22,23\rbrack} &
\multicolumn{1}{|l|}{$2.2246$} & $5.45$ & $0.919$ & $0.89024$ & $0.02876$ & $%
3.230$ \\ \hline
17. & \multicolumn{1}{|l|}{M\ \lbrack 24\rbrack} & \multicolumn{1}{|l|}{$%
2.22469$} & $5.424$ & $0.9043$ & $0.8845$ & $0.01975$ & $2.233$ \\ \hline
18. & \multicolumn{1}{|l|}{SCSS\ \lbrack 17\rbrack} & \multicolumn{1}{|l|}{$%
2.224575$} & $5.4193$ & $0.8838$ & $0.88300$ & $0.00079$ & $0.089$ \\ \hline
19. & \multicolumn{1}{|l|}{STS\ \lbrack 27\rbrack} & \multicolumn{1}{|l|}{$%
2.224575$} & $5.4194$ & $0.8845$ & $0.88303$ & $0.00147$ & $0.166$ \\ \hline
20. & \multicolumn{1}{|l|}{ER-C\ \lbrack 4\rbrack} & \multicolumn{1}{|l|}{$%
2.224575$} & $5.424$ & $0.8846$ & $0.88451$ & $0.00009$ & $0.011$ \\ \hline
21. & \multicolumn{1}{|l|}{KMMA\ \lbrack 8\rbrack} & \multicolumn{1}{|l|}{$%
2.224644$} & $5.412$ & $0.8883$ & $0.88191$ & $0.00639$ & $0.725$ \\ \hline
\end{tabular}
\end{center}

\newpage

\vspace{1pt}

\noindent \textbf{Table 2.} Parameters of the effective-range theory that
were calculated on the basis of the experimental values of $\varepsilon _{d}$%
, $A_{S}$, and $r_{d}$, as well as those found from the analysis of the
experimental $np$ scattering phase shifts

\begin{center}
\vspace{14pt}

\begin{tabular}{|c|c|c|c|c|c|}
\hline
Version & $a_{t},\,fm$ & $r_{t},\,fm$ & $P_{t}$ & $\rho _{m},\,fm$ & $\rho
_{d},\,fm$ \\ \hline
B & $5.4050$ & $1.75047$ & $-0.02312$ & $1.73716$ & $1.72385$ \\
S & $5.3863$ & $1.70257$ & $0.02010$ & $1.71321$ & $1.72385$ \\
K & $5.3713$ & $1.66391$ & $0.06064$ & $1.69388$ & $1.72385$ \\
ER & $5.395$ & $1.72385$ & $0$ & $1.72385$ & $1.72385$ \\
PWA & $5.4030$ & $1.7495$ & $-0.02592$ & $1.73461$ & $1.7197$ \\ \hline
\end{tabular}
\end{center}

\newpage

\vspace{1pt}

\begin{center}
\vspace{1pt}
\end{center}

\noindent \textbf{Table 3.} Triplet phase shift calculated for $np$
scattering in the shape-parameter approximation (28) with the parameter
values from Table 2 as a function of the laboratory energy $E_{lab}$

\begin{center}
\vspace{14pt}

\begin{tabular}{|c|cccccc|}
\hline
$E_{lab}$\thinspace , &  & Phase & shift & $\delta _{t}\,,$ & deg &  \\
\cline{2-2}\cline{2-7}\cline{3-7}
$MeV$ & Experiment \lbrack 38\rbrack & \multicolumn{1}{|c}{PWA} &
\multicolumn{1}{|c}{B} & \multicolumn{1}{|c}{S} & \multicolumn{1}{|c|}{K} &
\multicolumn{1}{|c|}{ER} \\ \hline
$0.1$ & $169.32$ & \multicolumn{1}{|c}{$169.315$} & \multicolumn{1}{|c}{$%
169.311$} & \multicolumn{1}{|c}{$169.349$} & \multicolumn{1}{|c|}{$169.380$}
& \multicolumn{1}{|c|}{$169.331$} \\
$0.5$ & $156.63$ & \multicolumn{1}{|c}{$156.645$} & \multicolumn{1}{|c}{$%
156.637$} & \multicolumn{1}{|c}{$156.728$} & \multicolumn{1}{|c|}{$156.802$}
& \multicolumn{1}{|c|}{$156.686$} \\
$1.0$ & $147.83$ & \multicolumn{1}{|c}{$147.823$} & \multicolumn{1}{|c}{$%
147.812$} & \multicolumn{1}{|c}{$147.954$} & \multicolumn{1}{|c|}{$148.068$}
& \multicolumn{1}{|c|}{$147.889$} \\
$2.0$ & $136.56$ & \multicolumn{1}{|c}{$136.548$} & \multicolumn{1}{|c}{$%
136.536$} & \multicolumn{1}{|c}{$136.770$} & \multicolumn{1}{|c|}{$136.958$}
& \multicolumn{1}{|c|}{$136.664$} \\
$5.0$ & $118.23$ & \multicolumn{1}{|c}{$118.236$} & \multicolumn{1}{|c}{$%
118.228$} & \multicolumn{1}{|c}{$118.750$} & \multicolumn{1}{|c|}{$119.168$}
& \multicolumn{1}{|c|}{$118.516$} \\
$10.0$ & $102.55$ & \multicolumn{1}{|c}{$102.598$} & \multicolumn{1}{|c}{$%
102.612$} & \multicolumn{1}{|c}{$103.672$} & \multicolumn{1}{|c|}{$104.521$}
& \multicolumn{1}{|c|}{$103.200$} \\
$20.0$ & $85.84$ & \multicolumn{1}{|c}{$86.049$} & \multicolumn{1}{|c}{$%
86.127$} & \multicolumn{1}{|c}{$88.385$} & \multicolumn{1}{|c|}{$90.211$} &
\multicolumn{1}{|c|}{$87.379$} \\
$30.0$ & $75.61$ & \multicolumn{1}{|c}{$76.043$} & \multicolumn{1}{|c}{$%
76.195$} & \multicolumn{1}{|c}{$79.702$} & \multicolumn{1}{|c|}{$82.592$} &
\multicolumn{1}{|c|}{$78.131$} \\
$40.0$ & $68.11$ & \multicolumn{1}{|c}{$68.822$} & \multicolumn{1}{|c}{$%
69.046$} & \multicolumn{1}{|c}{$73.796$} & \multicolumn{1}{|c|}{$77.806$} &
\multicolumn{1}{|c|}{$71.652$} \\
$45.0$ & $64.96$ & \multicolumn{1}{|c}{$65.843$} & \multicolumn{1}{|c}{$%
66.103$} & \multicolumn{1}{|c}{$71.462$} & \multicolumn{1}{|c|}{$76.050$} &
\multicolumn{1}{|c|}{$69.032$} \\
$50.0$ & $62.11$ & \multicolumn{1}{|c}{$63.171$} & \multicolumn{1}{|c}{$%
63.466$} & \multicolumn{1}{|c}{$69.424$} & \multicolumn{1}{|c|}{$74.602$} &
\multicolumn{1}{|c|}{$66.709$} \\
$100.0$ & $42.34$ & \multicolumn{1}{|c}{$45.705$} & \multicolumn{1}{|c}{$%
46.265$} & \multicolumn{1}{|c}{$57.612$} & \multicolumn{1}{|c|}{$69.487$} &
\multicolumn{1}{|c|}{$52.131$} \\ \hline
\end{tabular}
\end{center}

\newpage

\vspace{1pt}

\noindent \textbf{Table 4.} Difference of the experimental values of the
phase shift and its theoretical values calculated in the shape-parameter
approximation (28) with the parameter values from Table 2

\begin{center}
\vspace{14pt}

\begin{tabular}{|c|ccccc|}
\hline
$E_{lab}$\thinspace , &  &  & $\Delta \,,\,\deg $ &  &  \\
\cline{2-6}\cline{3-3}\cline{4-4}\cline{6-6}
$MeV$ & PWA & \multicolumn{1}{|c}{B} & \multicolumn{1}{|c}{S} &
\multicolumn{1}{|c|}{K} & \multicolumn{1}{|c|}{ER} \\ \hline
$0.1$ & $0.005$ & \multicolumn{1}{|c}{$0.009$} & \multicolumn{1}{|c}{$-0.029$%
} & \multicolumn{1}{|c|}{$-0.060$} & \multicolumn{1}{|c|}{$-0.011$} \\
$0.5$ & $-0.015$ & \multicolumn{1}{|c}{$-0.007$} & \multicolumn{1}{|c}{$%
-0.098$} & \multicolumn{1}{|c}{$-0.172$} & \multicolumn{1}{|c|}{$-0.056$} \\
$1.0$ & $0.007$ & \multicolumn{1}{|c}{$0.018$} & \multicolumn{1}{|c}{$-0.124$%
} & \multicolumn{1}{|c}{$-0.238$} & \multicolumn{1}{|c|}{$-0.059$} \\
$2.0$ & $0.012$ & \multicolumn{1}{|c}{$0.024$} & \multicolumn{1}{|c}{$-0.210$%
} & \multicolumn{1}{|c}{$-0.398$} & \multicolumn{1}{|c|}{$-0.104$} \\
$5.0$ & $-0.006$ & \multicolumn{1}{|c}{$0.002$} & \multicolumn{1}{|c}{$%
-0.520 $} & \multicolumn{1}{|c}{$-0.938$} & \multicolumn{1}{|c|}{$-0.286$}
\\
$10.0$ & $-0.048$ & \multicolumn{1}{|c}{$-0.062$} & \multicolumn{1}{|c}{$%
-1.122$} & \multicolumn{1}{|c}{$-1.971$} & \multicolumn{1}{|c|}{$-0.650$} \\
$20.0$ & $-0.209$ & \multicolumn{1}{|c}{$-0.287$} & \multicolumn{1}{|c}{$%
-2.545$} & \multicolumn{1}{|c}{$-4.371$} & \multicolumn{1}{|c|}{$-1.539$} \\
$30.0$ & $-0.433$ & \multicolumn{1}{|c}{$-0.585$} & \multicolumn{1}{|c}{$%
-4.092$} & \multicolumn{1}{|c}{$-6.982$} & \multicolumn{1}{|c|}{$-2.521$} \\
$40.0$ & $-0.712$ & \multicolumn{1}{|c}{$-0.936$} & \multicolumn{1}{|c}{$%
-5.686$} & \multicolumn{1}{|c}{$-9.696$} & \multicolumn{1}{|c|}{$-3.542$} \\
$45.0$ & $-0.883$ & \multicolumn{1}{|c}{$-1.143$} & \multicolumn{1}{|c}{$%
-6.502$} & \multicolumn{1}{|c}{$-11.090$} & \multicolumn{1}{|c|}{$-4.072$}
\\
$50.0$ & $-1.061$ & \multicolumn{1}{|c}{$-1.356$} & \multicolumn{1}{|c}{$%
-7.314$} & \multicolumn{1}{|c}{$-12.492$} & \multicolumn{1}{|c|}{$-4.599$}
\\
$100.0$ & $-3.365$ & \multicolumn{1}{|c}{$-3.925$} & \multicolumn{1}{|c}{$%
-15.272$} & \multicolumn{1}{|c|}{$-27.147$} & \multicolumn{1}{|c|}{$-9.791$}
\\ \hline
\end{tabular}
\end{center}

\newpage

\vspace*{2cm}

\begin{center}
\unitlength=0.24pt
\begin{picture}(1863,1192)
\put(0,1192){\special{em:graph figure.tif}}
\end{picture}
\end{center}

\vspace{1pt}

\noindent Correlation relationship between the dimensionless asymptotic
normalization constant for the deuteron, $C_{d}$, and the triplet scattering
length $a_{t}$. Points represent values obtained within various models of
nucleon-nucleon interaction in Table 1 (the numerals in the figure
correspond to the numbers of the lines in this table), while the straight
line was calculated by the approximate formula (21).

\end{document}